\documentclass{PoS}

\title{
A first prediction of the electromagnetic rare decays
$\eta^\prime\to\pi^0\gamma\gamma$ and $\eta^\prime\to\eta\gamma\gamma$
}

\ShortTitle{
A first prediction of the electromagnetic rare decays
$\eta^\prime\to\pi^0\gamma\gamma$ and $\eta^\prime\to\eta\gamma\gamma$
}

\author{
\speaker{Rafel Escribano}
\thanks{
I would like to express my gratitude to the QNP2012 Organizing Committee
for the opportunity of presenting this contribution, and for the pleasant and interesting workshop
we have enjoyed.}\\
Grup de F\'{\i}sica Te\`orica (Departament de F\'{\i}sica) and
Institut de F\'{\i}sica d'Altes Energies (IFAE),
Universitat Aut\`onoma de Barcelona,
E-08193 Bellaterra (Barcelona), Spain.
\\
E-mail: \email{rescriba@ifae.es}
}

\abstract{
The branching ratio of the electromagnetic rare decays $\eta\to\pi^0\gamma\gamma$ and $\eta^\prime\to (\pi^0,\eta)\gamma\gamma$
are ana\-lysed in terms of scalar and vector meson exchange contributions using the frameworks of the Linear Sigma Model and Vector Meson Dominance, respectively.
The measured $\eta\to\pi^0\gamma\gamma$ process serves as a test of our approach
while the non yet measured $\eta^\prime\to (\pi^0,\eta)\gamma\gamma$ reactions are predicted for the first time.
Our prediction for the $\eta\to\pi^0\gamma\gamma$ decay agrees with recent experimental reported values, thus supporting the validity of our framework.
Therefore, our predictions for the $\eta^\prime\to\pi^0\gamma\gamma$ and $\eta^\prime\to\eta\gamma\gamma$ decays
should be taken as a first indication of the possible values of the associated branching ratios.
We hope these predictions to be interesting and useful for experiments such as KLOE-2, Crystal Ball, WASA, and BES-III
where these processes are expected to be measured in the next future.
}

\FullConference{
Sixth International Conference on Quarks and Nuclear Physics,\\
April 16-20, 2012\\
Ecole Polytechnique, Palaiseau, Paris}

\begin{document}

\section{Introduction}
The electromagnetic rare decays
$\eta^\prime\to\pi^0\gamma\gamma$ and $\eta^\prime\to\eta\gamma\gamma$
are calculated for the first time with the aim of completing existing calculations on the related 
$\eta\to\pi^0\gamma\gamma$ process.
The first two, which have not yet been observed, will be potentially measured 
at several experiments such as KLOE-2, Crystal Ball, WASA, and BES-III
\cite{AmelinoCamelia:2010me,Li:2009jd},
while the latter, for which a measurement of its branching ratio and invariant mass spectrum
already exists \cite{Prakhov:2008zz}, will be certainly determined with higher precision.

From the experimental point of view, the situation at present is the following.
The branching ratio (BR) of $\eta\to\pi^0\gamma\gamma$ has been measured by
GAMS-2000 \cite{Alde:1984wj}, $\mbox{BR}=(7.2\pm 1.4)\times 10^{-4}$,
and CrystalBall@AGS in 2005 \cite{Prakhov:2005vx}, $\mbox{BR}=(3.5\pm 0.7\pm 0.6)\times 10^{-4}$,
and 2008 \cite{Prakhov:2008zz}, $\mbox{BR}=(2.21\pm 0.24\pm 0.47)\times 10^{-4}$,
the latter also including an invariant-mass spectrum for the two photons.
The PDG 2010 fit is $\mbox{BR}=(2.7\pm 0.5)\times 10^{-4}$ \cite{Nakamura:2010zzi}.
More recently, preliminary results from CrystalBall@MAMI \cite{Unverzagt:2009vm},
$\mbox{BR}=(2.25\pm 0.46\pm 0.17)\times 10^{-4}$,
and KLOE \cite{Gauzzi:2012zz,Di Micco:2005rv}, $\mbox{BR}=(0.84\pm 0.27\pm 0.14)\times 10^{-4}$,
have been reported as well.
For the $\eta^\prime\to\pi^0\gamma\gamma$ decay,
 only an upper bound exists, $\mbox{BR}<8 \times 10^{-4}$ at 90\% CL,
 obtained by the GAMS-2000 experiment \cite{Alde:1987jt} 25 years ago.
 Finally, for $\eta^\prime\to\eta\gamma\gamma$ there is no experimental evidence so far.
On the theory side, the $\eta\to\pi^0\gamma\gamma$ process
has been studied in many different frameworks,
Chiral Perturbation Theory (ChPT) \cite{Ametller:1991dp},
constituent quark model \cite{Ng:1993sc},
three-flavor Nambu-Jona-Lasinio model \cite{Nemoto:1996bh},
a chiral unitary approach \cite{Oset:2002sh,Oset:2008hp},
most of them in combination with the Vector Meson Dominance (VMD) prediction.
On the contrary, there are no theoretical analyses neither for
$\eta^\prime\to\pi^0\gamma\gamma$ nor for $\eta^\prime\to\eta\gamma\gamma$.

It is the purpose here to give an estimate of the
branching ratio of these three processes.
Since we are more interested in an estimate rather than a detailed calculation,
we will include in our analysis only the two main contributions,
that is, the exchange of an intermediate vector meson through the decay chain
$P^0\to V\gamma\to P^0\gamma\gamma$ plus the chiral loops.
Later, the chiral-loop prediction will be substituted by a Linear Sigma Model (L$\sigma$M)
calculation where the effects of scalar meson resonances are taken into account explicitly.
As a check of our approach, we first calculate these two contributions for the case of
$\eta\to\pi^0\gamma\gamma$.
Then, for the first time, we perform the same analysis for
$\eta^\prime\to\pi^0\gamma\gamma$ and $\eta^\prime\to\eta\gamma\gamma$.

\section{Chiral-loop prediction}
In ChPT, the tree-level contributions at ${\cal O}(p^2)$ and ${\cal O}(p^4)$ vanish because the pseudoscalar mesons involved are neutral.
The first non-vanishing contribution to $\eta\to\pi^0\gamma\gamma$ comes at  ${\cal O}(p^4)$, 
either from loops involving kaons, largely suppressed due to the kaon masses, 
or from pion loops, again suppressed since they violate G parity and are thus proportional to
$m_u-m_d$.
Numerically, it is seen to be three times smaller \cite{Ametller:1991dp}.
To simplify, we neglect the second contribution and work in the isospin limit.
The first sizable contribution comes at ${\cal O}(p^6)$,
but the coefficients involved are not well determined and one must resort to phenomenological models to fix them.
In this sense, for instance, VMD has been used to determine these coefficients by expanding the vector meson propagators and retain the lowest term.
This leads to values for the $\eta\to\pi^0\gamma\gamma$ decay rate
two times smaller than the ``all order'' estimate keeping the full vector meson propagator \cite{Ametller:1991dp}.
For the same process, the contributions of the scalar $a_0(980)$ and tensor $a_2(1320)$ resonances
to the ${\cal O}(p^6)$ chiral coefficients were calculated in the same manner
but no ``all order'' estimate was given in any case.
In addition, contrary to the VMD contribution where the coupling constant appears squared,
the signs of the $a_0$ and $a_2$ contributions are not unambiguously fixed.
On general grounds, one would expect the $a_2$ effects to be smaller than the $a_0$ ones
due to the heavier mass involved in the propagators.
For this reason, we will consider only the scalar meson contributions to the processes under analysis and
provide an ``all order'' estimate of these scalar effects based on a calculation performed in the L$\sigma$M model.
In this way, we will be able, first, to fix the sign ambiguity and, second, to test the relevance of including the full scalar meson propagators,
in a given model, instead of integrating them out. 
However, for the sake of completeness, we start considering the dominant chiral-loop contribution,
that is, the contributions containing two vertices of the lowest order Lagrangian and a charged pion or kaon loop.
The ${\cal O}(p^8)$ loop corrections from diagrams with two anomalous vertices are seen to be very small \cite{Ametller:1991dp}
and thus not considered here.
The explicit contributions of intermediate vector and scalar mesons are postponed to the next sections.

We start discussing the $\eta\to\pi^0\gamma\gamma$ case.
As stated before, the contribution from kaon loops is dominant and the pion loops vanish in the isospin limit.
The amplitude is written as
\begin{equation}
\label{Achietapi0}
{\cal A}^\chi_{\eta\to\pi^0\gamma\gamma}=
\frac{2\alpha}{\pi}\frac{1}{m_{K^+}^2}L(s_K)\{a\}\times{\cal A}^\chi_{K^+K^-\to\pi^0\eta}\ ,
\end{equation}
where
$\{a\}=(\epsilon_1\cdot\epsilon_2)(q_1\cdot q_2)-(\epsilon_1\cdot q_2)(\epsilon_2\cdot q_1)$,
$\epsilon_{1,2}$ and $q_{1,2}$ are the polarization and four-momentum vectors of the final photons,
$s_K=s/m_{K^+}^2$, $s=(q_1+q_2)^2=2q_1\cdot q_2$
is the invariant mass of the two photons,
$L(\hat s)$ is the loop integral defined as
\begin{equation}
\label{L}
L(z)=-\frac{1}{2z}-\frac{2}{z^2}f\left(\frac{1}{z}\right)\ ,\quad
f(z)=\left\{
\begin{array}{ll}
\frac{1}{4}\left(\log\frac{1+\sqrt{1-4z}}{1-\sqrt{1-4z}}-i\pi\right)^2 & \mbox{for}\ z<\frac{1}{4}\\[1ex]
-\left[\arcsin\left(\frac{1}{2\sqrt{z}}\right)\right]^2 & \mbox{for}\ z>\frac{1}{4}
\end{array}
\right.\ ,
\end{equation}
and ${\cal A}^\chi_{K^+K^-\to\pi^0\eta}$ is the four-pseudoscalar amplitude
\begin{eqnarray}
\label{AChPTKpKpi0eta}
{\cal A}^\chi_{K^+K^-\to\pi^0\eta}&=&\frac{1}{4f_\pi^2}
\left[\left(s-\frac{m_\eta^2}{3}-\frac{8m_K^2}{9}-\frac{m_\pi^2}{9}\right)(\cos\varphi_P+\sqrt{2}\sin\varphi_P)\right.\nonumber\\
&&\left.+\frac{4}{9}(2m_K^2+m_\pi^2)\left(\cos\varphi_P-\frac{\sin\varphi_P}{\sqrt{2}}\right)\right]\ ,
\end{eqnarray}
with $\varphi_P$ the $\eta$-$\eta^\prime$ mixing angle in the quark-flavour basis,
resulting from the loop computation
(not to confuse it with the four-pseudoscalar scattering amplitude calculated in ChPT at lowest order).
It is important to notice that in the seminal work of Ref.~\cite{Ametller:1991dp}
this chiral-loop prediction was computed taking into account the $\eta_8$ contribution alone and the mixing angle was fixed to
$\theta_P=\varphi_P-\arctan\sqrt{2}=\arcsin(-1/3)\simeq -19.5^\circ$.
Now, the $\eta_0$ contribution is also considered (in the large-$N_c$ limit where the pseudoscalar singlet is the ninth pseudo-Goldstone boson)
and the dependence on the mixing angle is made explicit.

For the $\eta^\prime\to\pi^0\gamma\gamma$ case,
the associated amplitude is that of Eq.~(\ref{Achietapi0}) but replacing $m_\eta\to m_{\eta^\prime}$,
$(\cos\varphi_P+\sqrt{2}\sin\varphi_P)\to (\sin\varphi_P-\sqrt{2}\cos\varphi_P)$ and
$(\cos\varphi_P-\sin\varphi_P/\sqrt{2})\to (\sin\varphi_P+\cos\varphi_P/\sqrt{2})$ in Eq.~(\ref{AChPTKpKpi0eta}).

Finally, for the $\eta^\prime\to\eta\gamma\gamma$ case, two amplitudes contribute,
one through a loop of charged kaons, as in the former two cases, and the other through a loop of charged pions, which in this case is not suppressed by $G$-parity.
Again, the corresponding amplitudes are that of Eq.~(\ref{Achietapi0}), replacing $s_K\to s_\pi$ and $m_{K^+}\to m_{\pi^+}$ for the pion loop,
with
\begin{eqnarray}
\label{AChPTKpKetaetap}
{\cal A}^\chi_{K^+K^-\to\eta\eta^\prime}&=&-\frac{1}{4f_\pi^2}
\left[\left(s-\frac{m_\eta^2+m_{\eta^\prime}^2}{3}-\frac{8m_K^2}{9}-\frac{2m_\pi^2}{9}\right)
\left(\sqrt{2}\cos2\varphi_P+\frac{\sin2\varphi_P}{2}\right)\right.\nonumber\\
&&\left.+\frac{4}{9}(2m_K^2-m_\pi^2)\left(2\sin2\varphi_P-\frac{\cos2\varphi_P}{\sqrt{2}}\right)\right]\ ,
\end{eqnarray}
and 
\begin{equation}
\label{AChPTpippimetaetap}
{\cal A}^\chi_{\pi^+\pi^-\to\eta\eta^\prime}=\frac{m_\pi^2}{2f_\pi^2}\sin2\varphi_P\ .
\end{equation}
The latter amplitude coincides with that of $\eta^\prime\to\eta\pi^+\pi^-$ when computed in the large-$N_c$ ChPT at lowest order
\cite{Escribano:2010wt}.
Needless to say, the former amplitudes for $\eta^\prime\to\pi^0\gamma\gamma$ and $\eta^\prime\to\eta\gamma\gamma$
constitute the first chiral-loop predictions of these two processes.

\section{VMD prediction}
Next to the chiral-loop amplitudes, there are also ``all order'' estimates of the corresponding exchange of intermediate vector bosons which are calculated in the framework of VMD.
The full VMD amplitude was seen to produce the dominant contribution to $\eta\to\pi^0\gamma\gamma$ \cite{Ametller:1991dp},
and the same happens, as we see below, for $\eta^\prime\to\pi^0\gamma\gamma$ and $\eta^\prime\to\eta\gamma\gamma$.
Now, we review the calculation for the $\eta\to\pi^0\gamma\gamma$ case, with some improvements with respect to Ref.~\cite{Ametller:1991dp},
and then calculate for the first time the full VMD amplitudes of $\eta^\prime\to\pi^0\gamma\gamma$ and $\eta^\prime\to\eta\gamma\gamma$.
For $\eta\to\pi^0\gamma\gamma$, the amplitude is written as
\begin{equation}
\label{AVMDetapi0}
{\cal A}^{\rm VMD}_{\eta\to\pi^0\gamma\gamma}=
\sum_{V=\rho, \omega, \phi}g_{V\eta\gamma}g_{V\pi^0\gamma}
\left[\frac{(P\cdot q_2-m_\eta^2)\{a\}-\{b\}}{D_V(t)}+
\left\{
\begin{array}{c}
q_2\leftrightarrow q_1\\
t\leftrightarrow u
\end{array}
\right\}\right]\ ,
\end{equation}
where
$t,u=(P-q_{2,1})^2=m_\eta^2-2P\cdot q_{2,1}$,
$\{b\}=(\epsilon_1\cdot q_2)(\epsilon_2\cdot P)(P\cdot q_1)+(\epsilon_2\cdot q_1)(\epsilon_1\cdot P)(P\cdot q_2)
-(\epsilon_1\cdot\epsilon_2)(P\cdot q_1)(P\cdot q_2)-(\epsilon_1\cdot P)(\epsilon_2\cdot P)(q_1\cdot q_2)$
and $D_V(t)=m_V^2-t-i\,m_V\Gamma_V$ are the vector meson propagators for $V=\omega, \phi$.
For the $\rho$ we use instead an energy-dependent
$\Gamma_\rho(t)=\Gamma_\rho\times[(t-4m_\pi^2)/(m_\rho^2-4m_\pi^2)]^{3/2}\times\theta(t-4m_\pi^2)$.
For $\eta^\prime\to\pi^0\gamma\gamma$ and $\eta^\prime\to\eta\gamma\gamma$,
the related amplitudes are Eq.~(\ref{AVMDetapi0}) with the replacements
$g_{V\eta\gamma}g_{V\pi^0\gamma}\to g_{V\eta^\prime\gamma}g_{V\pi^0\gamma}$ and
$g_{V\eta\gamma}g_{V\pi^0\gamma}\to g_{V\eta^\prime\gamma}g_{V\eta\gamma}$, respectively,
and $m_\eta^2\to m_{\eta^\prime}^2$.
When the OZI-rule is applied, that is $\omega=(u\bar u+d\bar d)/\sqrt{2}$ and $\phi=s\bar s$,
the corresponding couplings are
\begin{equation}
\label{VMDcouplings}
\begin{array}{c}
\begin{array}{ll}
g_{\rho\eta\gamma}g_{\rho\pi^0\gamma}=
g_{\omega\eta\gamma}g_{\omega\pi^0\gamma}=\left(\frac{Ge}{\sqrt{2}g}\right)^2\frac{1}{3}\cos\varphi_P\ ,\quad
& g_{\phi\eta\gamma}g_{\phi\pi^0\gamma}=0\ ,\\[2ex]
g_{\rho\eta^\prime\gamma}g_{\rho\pi^0\gamma}=
g_{\omega\eta^\prime\gamma}g_{\omega\pi^0\gamma}=\left(\frac{Ge}{\sqrt{2}g}\right)^2\frac{1}{3}\sin\varphi_P\ ,\quad
& g_{\phi\eta^\prime\gamma}g_{\phi\pi^0\gamma}=0\ ,\\[2ex]
\end{array}\\
g_{\rho\eta^\prime\gamma}g_{\rho\pi^0\gamma}=
9g_{\omega\eta^\prime\gamma}g_{\omega\pi^0\gamma}=
-\frac{9}{4}g_{\phi\eta^\prime\gamma}g_{\phi\pi^0\gamma}=
\left(\frac{Ge}{\sqrt{2}g}\right)^2\cos\varphi_P\sin\varphi_P\ ,
\end{array}
\end{equation}
where $G=3g^2/(4\pi^2 f_\pi)$ and $g$ is the vector-pseudoscalar-pseudoscalar coupling constant of VMD
which can be fixed from various $\rho$ and $\omega$ decay data.

In Ref.~\cite{Ametller:1991dp}, the VMD prediction for $\eta\to\pi^0\gamma\gamma$
was calculated assuming equal $\rho$ and $\omega$ contributions and without including the decay widths in the propagators.
In this case, these approximations are valid since the phase space available prevents the vector mesons to resonate.
However, for $\eta^\prime\to\pi^0\gamma\gamma$,
the phase space allowed permits these vectors to be on-shell and the introduction of their decay widths is mandatory.
For this reason, we include, for all the three cases, the decay widths in the vector meson propagators.

\section{L$\sigma$M prediction}
An "all order" estimate of the scalar meson exchange effects to the processes under study can be achieved in the L$\sigma$M
where the complementarity between this model and ChPT can be used to include the scalar meson poles
at the same time as keeping the correct low-energy behavior expected from chiral symmetry.
This procedure was applied with success to the related $V\to P^0 P^0\gamma$ decays
\cite{Escribano:2006mb}.
The $a_0(980)$ enters into the calculation of $\eta\to\pi^0\gamma\gamma$ and $\eta^\prime\to\pi^0\gamma\gamma$,
more intensively in the latter case on account of phase space,
while the $\sigma(600)$ and $f_0(980)$ do the same in $\eta^\prime\to\eta\gamma\gamma$,
although only the first contributes in a substantial way.
Taking into account the scalar meson effects in an explicitly way does not provide a noticeable improvement
with respect to the chiral-loop prediction, except for the case of $\eta^\prime\to\eta\gamma\gamma$
where the $\sigma$ contribution turns out to be considerable.
However, the details of this calculation are involved and will be described elsewhere \cite{Escribano&Jora}.

\section{Preliminary results}
\begin{table}
\begin{center}
\begin{tabular}{|l|c|c|c|c|c|c|}
\hline
& chiral loops & L$\sigma$M & VMD & $\Gamma$ & BR$_{\rm th}\times 10^4$ & BR$_{\rm exp}\times 10^4$\\
\hline
$\eta\to\pi^0\gamma\gamma$\hfill (eV) & $1.24\times 10^{-3}$ & $4.5\times 10^{-4}$ & 0.26 & 0.28 & 2.1 & $2.7\pm 0.5$\\
\hline
$\eta^\prime\to\pi^0\gamma\gamma$\ (keV) & $7.7\times 10^{-5}$ & $1.3\times 10^{-4}$ & 1.29 & 1.29 & 65 & <8\ (90\% CL)\\
\hline
$\eta^\prime\to\eta\gamma\gamma$\hfill (eV) & $1.4\times 10^{-2}$ & 0.96 & 48.8 & 51.2 & 2.6 & ---\\
\hline
\end{tabular}
\end{center}
\caption{
Chiral-loop, L$\sigma$M and VMD predictions for  $\eta\to\pi^0\gamma\gamma$, $\eta^\prime\to\pi^0\gamma\gamma$
and $\eta^\prime\to\eta\gamma\gamma$.
The total decay widths are calculated from the coherent sum of the L$\sigma$M and VMD contributions.
The comparison between the predicted branching ratios and the present experimental values, if available, is also performed.}
\label{table1}
\end{table}

The preliminary results of our analysis are shown in Table \ref{table1},
where the predictions of chiral loops, the L$\sigma$M, which replaces the former when scalar meson poles are incorporated,
VMD, and the total decay width and branching ratio for the three processes are included.
The comparison with the experimental results, if available, is also displayed.
The total decay width is the result of adding the L$\sigma$M and VMD contributions coherently.
For the numerical results, we use $f_\pi=92.2$ MeV, $|g|=4.2$ from the present value of $\rho\to\pi\pi$,
and $\varphi_P=(40.4\pm 0.6)^\circ$ \cite{Ambrosino:2009sc} for the $\eta$-$\eta^\prime$ mixing angle.
For $\eta\to\pi^0\gamma\gamma$,
our calculation agrees with the ``all-order estimate'' of Ref.~\cite{Ametller:1991dp}
and the more involved analysis of Refs.~\cite{Oset:2002sh,Oset:2008hp},
thus giving support to our approach as a starting point for the determination of the other two processes.
For $\eta^\prime\to\pi^0\gamma\gamma$, the intermediate vector meson contributions dominate and the scalar meson effects 
are seen to be negligible.
The $\omega$ contribution prevails with a 80.2\% of the total VMD signal, while the $\rho$ contributes with a 4.6\%.
The predicted branching ratio appears to be one order of magnitude bigger than the old experimental upper bound.
Therefore, a new measurement would be welcome.
Finally, for $\eta^\prime\to\eta\gamma\gamma$,
the VMD contribution also dominates but the scalar meson effects seem to be sizable, in particular those related with the $\sigma$ meson.
The interference term is constructive.
The $\rho$, $\omega$ and $\phi$ contribute with a 59.9\%, 15.8\% and 1.6\%, respectively,
while the L$\sigma$M calculation enhances by two orders of magnitude the chiral-loop prediction.
Since $G$-parity does not apply to this case, the loop of charged kaons is suppressed and only the charged-pion loop plays a role.

In summary, the calculated branching ratios for $\eta^\prime\to\pi^0\gamma\gamma$ and $\eta^\prime\to\eta\gamma\gamma$
give values which we consider are large enough to be measured in the near future by several experimental collaborations.

\section*{Disclaimer}
For reasons of limited space, we do not include any of the two-photon invariant mass spectra which will be accessible in Ref.~\cite{Escribano&Jora}.


\end{document}